\documentclass[sigconf]{acmart}
\AtBeginDocument{%
  }

\copyrightyear{2026}
\acmYear{2026}
\setcopyright{cc}
\setcctype{by-nc-nd}
\acmConference[ISLPED '26]{ACM/IEEE International Symposium on Low Power Electronics and Design}{August 05--07, 2026}{Evanston, IL, USA}
\acmBooktitle{ACM/IEEE International Symposium on Low Power Electronics and Design (ISLPED '26), August 05--07, 2026, Evanston, IL, USA}
\acmDOI{10.1145/3816440.3818565}
\acmISBN{979-8-4007-2748-1/2026/08}

\usepackage{pgfplots}
\usepackage{tikz}
\usepackage{xcolor}
\usepackage{glossaries}
\glsdisablehyper

\newacronym{rtc}{RTC}{Real-Time clock}
\newacronym{pmic}{PMIC}{Power Management Integrated Circuit}
\newacronym{ic}{IC}{Integrated Circuit}
\newacronym{mcu}{MCU}{microcontroller}
\newacronym{ssn}{SSN}{Smart Sensor Node}
\newacronym{iot}{IoT}{Internet of Things}
\newacronym{pv}{PV}{photovoltaic}
\newacronym{dpm}{DPM}{dynamic power management}
\newacronym{pm}{PM}{power management}
\newacronym{dvfs}{DVFS}{Dynamic Voltage and Frequency Scaling}
\newacronym{mlp}{MLP}{multilayer perceptron}
\newacronym{ble}{BLE}{Bluetooth Low Energy}
\newacronym{rf}{RF}{radio frequency}
\newacronym{pzt}{PZT}{piezoelectric}
\newacronym{dt}{DT}{decision tree}
\newacronym{soa}{SoA}{state-of-the-art}
\newacronym{ml}{ML}{Machine Learning}
\newacronym{aoout}{AOOUT}{Always-On Output}
\newacronym{lvout}{LVOUT}{Low-Voltage Output}
\newacronym{hvout}{HVOUT}{High-Voltage Output}

\begin{document}

\title{A Hardware-Based Multi-Stage Dynamic Power Management Architecture for Autonomous Low-Light Operation}

\author{Charalampos S. Kouzinopoulos}
% \email{charis.kouzinopoulos@maastrichtuniversity.nl}
\orcid{0000-0001-8829-504X}
\affiliation{%
  \institution{Maastricht University}
  \city{Maastricht}
  \country{The Netherlands}
}

\author{Marcel Louis Meli}
% \email{mema@zhaw.ch}
\orcid{0009-0001-7266-3028}
\affiliation{%
  \institution{ZHAW School of Engineering, Zurich University of Applied Sciences}
  \city{Zurich}
  \country{Switzerland}
}

\author{Martin Schellenberg}
\orcid{0009-0006-9304-6359}
% \email{martin.schellenberg@microdul.com}
\author{Philip J. Poole}
% \email{phil.poole@microdul.com}
\affiliation{%
  \institution{Microdul AG}
  \city{Zurich}
  \country{Switzerland}
}

\author{Mathieu Bellanger}
% \email{mathieu.bellanger@lightricity.co.uk}
\author{Matthias Kauer}
\orcid{0000-0002-5442-7574}
% \email{matthias.kauer@lightricity.co.uk}
\affiliation{%
  \institution{Lightricity Limited}
  \city{Oxford}
  \country{United Kingdom}
}

\author{Julien De Vos}
% \email{julien.devos@e-peas.com}
\affiliation{%
  \institution{e-peas S.A.}
  \city{Mont-Saint-Guibert}
  \country{Belgium}
}

\author{Dimosthenis Ioannidis}
% \email{djoannid@iti.gr}
\orcid{0000-0002-5747-2186}
\author{Dimitrios Tzovaras}
% \email{Dimitrios.Tzovaras@iti.gr}
\orcid{0000-0001-6915-6722}
\affiliation{%
  \institution{CERTH/ITI}
  \streetaddress{P.O.Box 60361, 6th km Harilaou - Thermi}
  \city{Thermi}
  \country{Greece}
  \postcode{57001}
}

% \affiliation{%

% \affiliation{%

% \affiliation{%

% \affiliation{%

% \affiliation{%

% \affiliation{%

% \affiliation{%

\renewcommand{\shortauthors}{C.S. Kouzinopoulos et al.}

\begin{abstract}
The advance of autonomous Smart Sensor Networks and embedded systems for the Internet of Things, powered by photovoltaic energy harvesting, is severely limited by energy efficiency, especially in low-light environments. While Dynamic Power Management is essential for energy conservation, conventional software-based techniques that rely on processor-managed low-power states incur a persistent quiescent current drain. This current becomes the dominant energy sink in energy-scarce conditions, limiting autonomy. The work of this paper addresses this limitation by introducing a robust, hardware-orchestrated dynamic power management architecture that improves existing configurations for battery-based sensor nodes. The proposed architecture achieves a minimal quiescent drain of $452nA$, by completely power-gating the microcontroller and all non-essential peripherals, with wake-up orchestrated by an ultra-low-power PMIC, RTC and a novel latch circuit developed specifically for this work. Our evaluation demonstrates that the dynamic power management architecture is significantly more efficient than traditional software-based sleep modes. 
% We validate this in a real-world case study, demonstrating a net-positive energy gain of $23.9mJ$ per $10$-minute cycle at an illuminance of $200$ lux.
\end{abstract}




% \received{20 February 2026}
% \received[revised]{12 March 2026}
% \received[accepted]{5 June 2026}

\maketitle

\section{Introduction}
The rapid development of \Gls*{iot} applications has led to the widespread deployment of \glspl*{ssn}, embedded, autonomous devices for data-driven automation at the edge. The nodes integrate sensors for data acquisition with wireless communication capabilities, while recent advances have enabled the integration of \textit{edge AI}; the execution of AI models at the edge~\cite{bröcheler2025segmentedrobotgraspingperception}\cite{kokhazadeh2025cnn}\cite{kroger2025device} for localized decision-making.
To achieve autonomy, these systems are often powered by energy harvesting sources such as \gls*{pv} elements. In such energy-constrained settings, particularly in indoor or variable outdoor environments, energy budget is largely dictated by low-light performance.

\Gls*{dpm}~\cite{de2017autonomous}\cite{devadas2010interplay}\cite{narang2023uncertainty} is a fundamental approach to maximize operational lifetime, typically involving \Gls*{dvfs}, clock gating, establishment of Voltage-Frequency Islands with dynamically adjustable Voltage/Frequency knobs, as well as event-driven sleep and low-power modes. However, these strategies share a fundamental limitation: they are typically software-orchestrated. The main system \gls*{mcu} must remain in a deep-sleep or standby state to manage wake-up events. This active-sleep state, while low-power, incurs a persistent quiescent current (indicatively $2.57\mu A$ for an STM32L4 \gls{mcu} in stop $2$ mode). In ultra-low-light conditions where harvested power is minimal, this quiescent drain becomes the dominant energy sink, creating an operational floor for autonomous operation.

To improve upon this gap, we build upon and extend recent work in battery-less systems~\cite{meli2020low2}\cite{meli2023energy} to propose a hardware-orchestrated \gls{dpm} architecture for battery-based sensor nodes that completely power-gates the \gls*{mcu} and most peripherals during sleep cycles, eliminating their quiescent current. This architecture is built around the always-on power domain of a \gls*{pmic}, powering only a low-power \gls*{rtc} and a capacitive touch sensor for periodic or on-demand state switching. These hardware components are responsible for orchestrating the wake-up and shutdown of the main compute domain.
% This paper also presents the design and implementation of a miniaturized, autonomous embedded system built around this architecture. The system integrates an $80MHz$ ARM Cortex-M4 \gls*{mcu}, multiple short- and long-range wireless \glspl*{ic} (BLE, LoRa, NFC) and sensors, powered by a $10mAh$ battery and a $2cm^2$ high-efficiency \gls*{pv} panel. The final prototype has a form factor of a credit card with a very slim profile. 

\textbf{Contributions.} The main contributions of this work are:

\begin{itemize}
    \item A robust hardware-orchestrated \Gls*{dpm} architecture for battery-based sensor nodes that eliminates \gls*{mcu} and peripheral quiescent current by fully power-gating the compute domain, achieving a measured sleep-state drain of $452nA$.
    \item A new latch circuit for the capacitive sensor, forming a key part of the \Gls*{dpm} architecture. This circuit, which enables the sensor to provide a persistent \textit{enable} signal from a momentary trigger, was designed by the authors and integrated into the commercial Microdul MS8892 capacitive sensor.
    \item The design of an autonomous, energy-harvesting \gls{ssn}, integrating the proposed \Gls*{dpm} architecture. 
    \item Experimental results demonstrating cold-starting capability and sustained, net-positive energy operation under very low illuminance levels.

    % down to \color{red}XXX\color{black} lux
    % Compared to similar systems in the literature employing a combination of PV harvester, battery and \Gls*{pmic} for energy storage and distribution, the proposed node achieves reduced quiescent current during inactive periods, improving energy efficiency and autonomy.

    % \item Evaluation of the complete, integrated system as a wearable on a personal thermal comfort scenario
\end{itemize}

% This paper is structured as follows:
%     Section~\ref{Section:Related} presents current research on integrated multi-sensor embedded systems.
%     Section~\ref{sec:power_management} provides a comnprehensive analysis of the proposed \gls*{dpm} architecture. 
%     Section~\ref{Section:Architecture} describes the design of an autonomous  \gls*{ssn}, integrating the \gls*{dpm}.
%     Section~\ref{Section:Evaluation} presents an extensive performance evaluation, including a real-world case study.
%     % as well as details on the experimental evaluation of the system for a thermal comfort scenario
%     Finally, section~\ref{Section:Conclusions} discusses conclusions and ideas for future work.

\section{Related Work}\label{Section:Related}
% This section summarizes the main relevant research on power management architectures for low-power, multi-sensor systems. 
% A summary of energy harvesting sources typically used in \glspl*{ssn} is also provided here.

% embedded, miniaturized, low-power, multi-sensor systems and SSNs that are powered by energy harvesting. 
% An extensive analysis of the \gls*{soa} research for IoT systems powered with solar cells as well as on harvesting architectures, \Gls*{rtc} and low-power timers and touch detection can be found in~\cite{meli2023energy}. 

% Additional works from the literature on \gls*{dpm} architectures for \glspl*{ssn} are presented in detail in section~\ref{Section:Comparison} below for the comparison of the proposed system with the \gls*{soa}. 

% Power management optimization for energy-harvesting autonomous \glspl*{ssn} has been extensively studied. 
% Such nodes are usually inactive the majority of time; they are activated only during specific time-windows, either periodically or as a response to an external event, perform measurements, processing and/or data transmission and subsequently return back to an inactive state.

Most prior work on power management optimization for energy-harvesting autonomous \glspl*{ssn} focuses on aggressive duty-cycling, where components enter idle, stop or deep sleep during periods of inactivity. While effective, these approaches still incur significant quiescent currents, limiting energy autonomy under low-light conditions (i.e.  \cite{bregar2022power}\cite{sidibe2022multifunctional}).
To further reduce leakages, several designs actively cut-off power to non-essential components during inactive periods (i.e. \cite{la2021energy}\cite{la2022battery}). However, in these approaches the \gls*{mcu} typically remains powered, reducing system autonomy.

Some works fully power down the main load, including the \gls*{mcu} and most non-essential peripherals, while keeping only a minimal set of ultra-low-power components active~ \cite{meli2020low2}\cite{meli2023energy}. These designs achieve significant energy efficiency under ultra-low-light conditions as low as $5$-$20$ lux. However, they are typically battery-less and rely exclusively on small capacitive storage, which severely limits the maximum supported load and edge processing capabilities.

% This section provides a comparative analysis between the proposed work of this paper and the aforementioned state of the art approaches. The results are summarized in Table~\ref{tab:SoA_comparison}.

For example, \cite{meli2023energy} presents a \Gls*{ble} node operating down to $5$ lux by duty-cycling between a $310$nA Deep Sleep state and a Normal mode either periodically using an \Gls*{rtc} or based on a user command via a capacitive touch sensor. 
Unlike our proposed architecture though, this system lacks a dedicated \Gls{pmic} and battery. Instead, it relies solely on a $100\mu F$ capacitor charged directly from \gls*{pv}, restricting its application to minimal, highly intermittent loads. 
% Driven by a $4$cm$^2$ \gls*{pv} array, the capacitor reaches a maximum of $3.15V$ with a variable load of $9500kOhm$ as its Maximum Power Point (MPP); under $19$ lux it is measured at $3.37V$ with a variable load of $2200kOhm$ as its MPP.
Although a custom ASIC \Gls*{pmic} is included, with a reported operating current of $30nA$ at room temperature, only a single comparator of the \gls*{ic} is utilized to monitor the voltage of the capacitor; the system can be powered up as long as the supercapacitor's voltage is at least $2.25V$. 
The quiescent current consumption of the system in Deep Sleep can be approximated to $140nA$ based on the consumption of the comparator, \Gls*{rtc} and touch sensor.
Similarly, \cite{meli2020low2} reports a LoRaWAN node operating down to $20$ lux using a dedicated \Gls*{pmic} and a $10mF$  supercapacitor to power a LoRa SX1276 transceiver, a BME680 sensor and a Cortex-M4 \gls*{mcu}, requiring $8mJ$ per $10$-minute active cycle.
The limits of the \Gls*{pmic} are set in such a way that the system will start operating as soon as the charge of the supercapacitor reaches $3V$ and it will continue charging up to a maximum of $3.6V$. After the $3V$ threshold is reached, the supercapacitor can be discharged down to $2V$. 
A total load of up to $44.8mJ$ can be supported by this configuration. A larger supercapacitor can be used to support increased loads, at the disadvantage of requiring additional time to charge it at similar illumination levels.

\begin{table}[h]
  \caption{State-of-the-art comparison}
  \label{tab:SoA_comparison}
  \begin{tabular}{ccc}
    \toprule
    Work & Always-on components & Always-on current\\
    \midrule
    \cite{meli2023energy} & \begin{tabular}{@{}c} MA198 (Comparator), \\ RV-3028-C7, MS8892\end{tabular} & \begin{tabular}{@{}c}$30nA + 45nA$\\ $+ \, 65nA$\end{tabular} \\
    \cite{meli2020low2} & EM8502 & -\\
    \cite{la2021energy}\cite{la2022battery} & STM32L0 & $1\mu A$\\
    \cite{bregar2022power} & nRF8001, MSP430F2274 & $2\mu A$ + $1\mu A$\\
    \cite{sidibe2022multifunctional} & \begin{tabular}{@{}c}QN9080, BQ22570,\\ AEM30940, HDC2080\end{tabular} & $27\mu A$\\
    \begin{tabular}{@{}c} \textbf{This}\\ \textbf{work}\end{tabular} &
    \begin{tabular}{@{}c}\textbf{\Gls*{pmic} based on AEM10941},\\ \textbf{RV-3028-C7, MS8892}\end{tabular} & \begin{tabular}{@{}c}$310nA$ (datasheet)\\ $452nA$ (measured)\end{tabular}\\
  \bottomrule
\end{tabular}
\end{table}

An autonomous \Gls*{ble} node is presented in~\cite{la2021energy}. The node uses the BLUENRG-2 \Gls*{ble} SoC as a load, together with an STM32L0-based \gls*{mcu} to control the transition between harvesting and data transmission phase. During the harvesting phase, an AM-1606C solar cell is used to harvest ambient energy down to $200$ lux to a $22\mu F$ capacitor. In low-power stop mode, the \gls*{mcu} stays powered-on, operating as a voltage detector to monitor the voltage across the capacitor.
As soon as the capacitor voltage reaches a limit of $3.3V$, the node transitions to the data transmission phase.
During that phase, the \gls*{mcu} switches to run mode while the \Gls*{ble} \gls*{ic} is utilized in advertising mode to transmit $7$-byte advertising data packets with an output power of $+8dBm$ over three channels.
The node returns to the harvesting phase based on an interrupt from the low-power timer of the \gls*{mcu}.
The quiescent current of the node during its harvesting phase is equal to the operating current of the \gls*{mcu} in stop mode, reported at approximately $1\mu A$.
A system with similar characteristics and the addition of the HTS221 \gls*{ic} for moisture and temperature sensing is detailed in~\cite{la2022battery}.

The autonomy of a wearable node for ECG monitoring is detailed in~\cite{bregar2022power}. As loads, the node uses the MSP430F2274 \gls*{mcu} for processing and ADC, the nRF8001 \Gls*{ble} \gls*{ic} for short-range communication, and an analog circuit for ECG measurements, including a pre-amplifier and analog filters. For energy storage, the system uses a $240mAh$ li-ion battery.
During operation, the node transitions between: a sleep phase where the load is set into low-power modes; an idle phase where the \Gls*{ble} \gls*{ic} performs advertising; and an active phase where the sensor performs ECG sampling.
During sleep mode, nRF8001 is put into idle phase with a reported quiescent current of $2\mu A$ while the \gls*{mcu} consumes approximately $1\mu A$ in low-power mode 3.

A sensor node for the monitoring of physical parameters in reinforced concrete is presented in~\cite{sidibe2022multifunctional}. The node is powered through wireless power transfer using RF to DC rectifiers to charge a $150\mu F$ capacitor, using two \Glspl*{pmic} for energy distribution, BQ22570 and AEM30940. The node transitions between a deep sleep, with an average current of $27\mu A$, and an advertising phase.

To summarize, existing work exhibits a fundamental trade-off between ultra-low quiescent battery-less architectures with limited load capability, and
battery-operated systems with higher always-on consumption (Table~\ref{tab:SoA_comparison}).
We improve the state of the art by integrating a near-zero quiescent drain ($452nA$) on battery-based systems, characteristic of battery-less designs, through a new hardware-orchestrated \Gls*{dpm} architecture together with a novel latch circuit for the capacitive sensor. 

\section{Multi-stage \gls*{dpm} architecture}\label{sec:power_management}
\begin{figure}[h]
	\centering		\includegraphics[width=\linewidth]{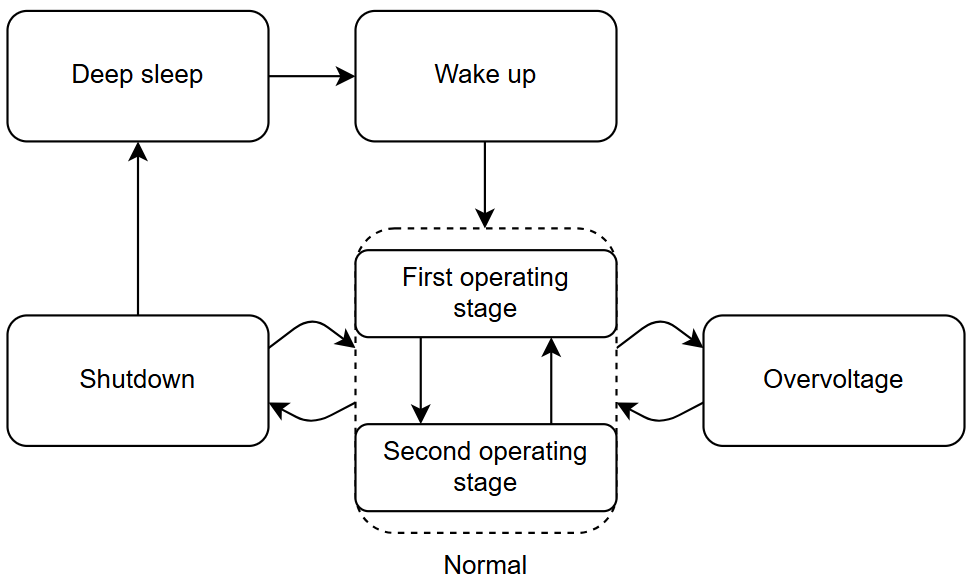}
	\caption{System modes}
	\label{Fig:PMIC2}
\end{figure}

In order to minimize energy consumption, we have developed a multi-stage \gls*{dpm} architecture that utilizes different operating modes, as depicted in Figure~\ref{Fig:PMIC2}: Deep sleep; Wake up; Normal; Overcharge and Shutdown modes. Switching between modes does not involve the use of the \gls*{mcu} to minimize power consumption.

\begin{itemize}
    \item In \textit{Deep sleep mode}, the energy storage is deeply discharged and there is no available energy to be harvested from the energy harvester
    \item When a cold-start voltage of $300mV$ and $2\mu W$ of power become available from the harvester, the \Gls*{pmic} enters \textit{Wake up mode}. If the voltage of the storage element is below a threshold $V_{CHRDY}$, it will first need to be charged through the \Gls*{pmic}
    \item When the voltage of the storage element reaches $V_{CHRDY}$, the \Gls*{pmic} enters \textit{Normal mode}; the outputs \gls{lvout} and \gls{aoout} are shorted to the buck regulator output and the external DC/DC converter can drive \gls{hvout}. During Normal mode, there can be three different scenarios depending on the voltage of the storage element:
    \begin{itemize}
        \item If the voltage reaches a higher threshold $V_{OVCH}$, the \Gls*{pmic} enters \textit{Overcharge mode}; the charging of the storage stops and its voltage is maintained to approximately $V_{OVCH}$ with hysteresis, to prevent damage from overcharging
        \item If the storage element's voltage ranges between $V_{CHRDY}$ and $V_{OVCH}$, the \Gls*{pmic} stays in Normal mode
        \item If its voltage drops below $V_{CHRDY}$, the \Gls*{pmic} enters \textit{Shutdown mode} to prevent deep discharge potentially leading to damage to the energy storage element. 
        If sufficient energy becomes available within approximately $600ms$, the \Gls*{pmic} returns to Normal mode. If not, the circuit returns to Deep sleep mode
    \end{itemize}
\end{itemize}

The Normal mode can be further divided into two operating stages, the first operating stage and the second operating stage\footnote{Note that the system can be easily extended to include additional operating stages}. In the first operating stage, only \gls{aoout} is active. No other components, including the \gls*{mcu}, are powered during this stage. The system is operational and can be woken-up, while maintaining the lowest possible power consumption.

For the system to wake up and enter the second operating stage, the \gls{lvout} power domain must be turned on. The \Gls*{pmic} provides an enable input, 2.2\_SW\_EN, which controls this power domain. The capacitive sensor and \Gls*{rtc} can trigger the enable input of the \Gls*{pmic} in order to turn on the \gls{lvout} switched domain and subsequently power on the system. The capacitive sensor can power the system based on a touch interaction from the user. The \Gls*{rtc} on the other hand is programmed during system runtime in order to generate periodical wake up triggers. The enable signal 2.2\_SW\_EN is latched in the capacitive sensor; it means that the power domain stays on when the enable event is revoked. This feature provides a trigger-style enabling of the $2.2V$ switched domain.
An optional, third \gls{hvout} switched domain is also provided by the \Gls*{pmic}, in order to power components that require a higher voltage such as CO$_2$ sensors. This domain is controlled via the \gls*{mcu}.

\begin{figure} [h]
	\centering		\includegraphics[width=\linewidth]{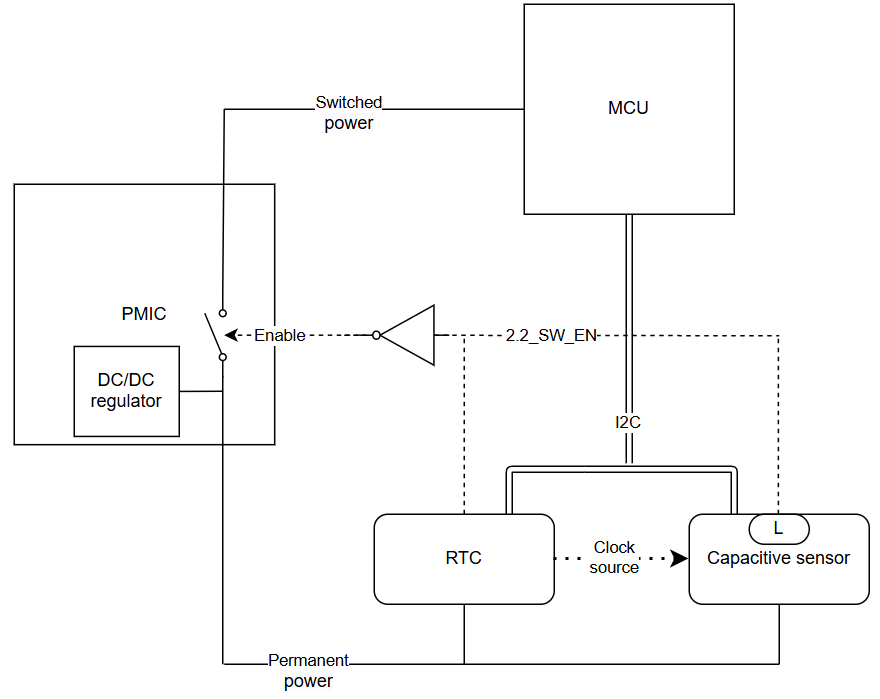}
	\caption{Switching between operating stages}
	\label{Fig:MS8892-2}
\end{figure}

A key challenge in the proposed hardware \gls{dpm} design is translating the momentary triggers from the RTC or capacitive sensor into a persistent \textit{enable} signal that holds the main compute domain \textit{on}.
Figure~\ref{Fig:MS8892-2} depicts the novel latching mechanism designed to address this problem. This circuit, a primary contribution of this work, is integrated with the capacitive sensor's logic to switch between operating stages.

The high level on signal 2.2\_SW\_EN is achieved by a pull-up resistor. If 2.2\_SW\_EN is driven to logic low level by MS8892, its internal pull-up resistor is disconnected to avoid the static current flow and further reduce power consumption.
The system power state is kept in the internal latch $L$. The latch $L$ will be set by a touch event or when the \Gls*{rtc} pulls the signal 2.2\_SW\_EN low, thus switching on the switched power via enable.
After power-up, the \gls*{mcu} is in control and can read out the wake-up source from the MS8892. It can re-configure the parameters of the MS8892 and the alarm settings of the \Gls*{rtc}.
The system will return to the deep sleep mode by clearing the latch in the MS8892, which in turn disables the switched power supply. The latch can be cleared by an \gls*{mcu} command via the I$^2$C bus or alternatively via a digital signal from the \gls*{mcu} to the MS8892 via 2.2\_SW\_DIS.
This design was developed and validated in collaboration with the component manufacturer, Microdul, and was adopted for their commercial MS8892 IC.

\begin{figure*}
  \centering
  \includegraphics[width=\linewidth]{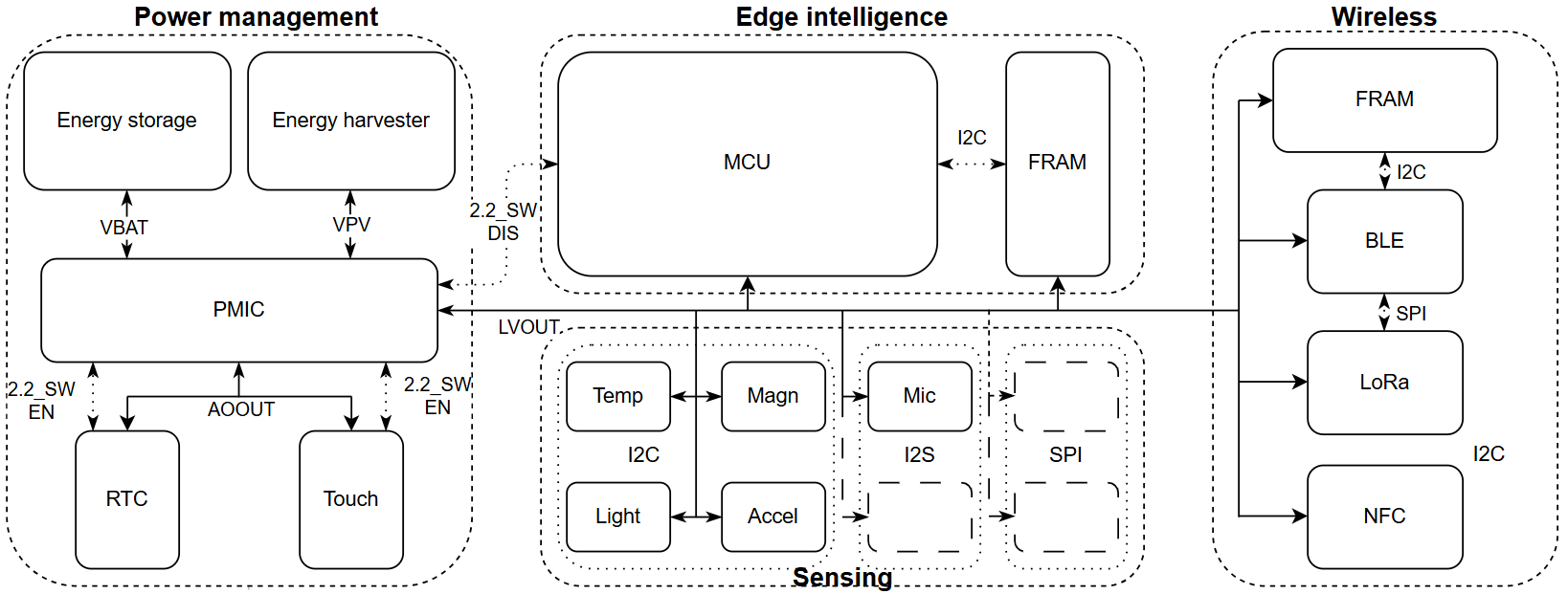}
  \caption{High-level architecture of the \gls{ssn}}
  \label{Fig:Architecture}
  \Description[Blocks of components]{A high-level architecture depicting the list of interconnected blocks and their main components.}
\end{figure*}

Since the interfacing components are always powered, it is critical that they consume minimal energy. For this reason, a single oscillator is permanently running for both the capacitive sensor and the \Gls*{rtc}. This is made possible by an external clock input to the capacitive sensor and using the oscillator of the \Gls*{rtc} to trigger the periodic capacitive touch detection.

\section{Sensor node implementation}\label{Section:Architecture}

\begin{figure}[h]
    \centering
    \begin{minipage}{0.43\textwidth}
        \centering
        \includegraphics[width=1\textwidth]{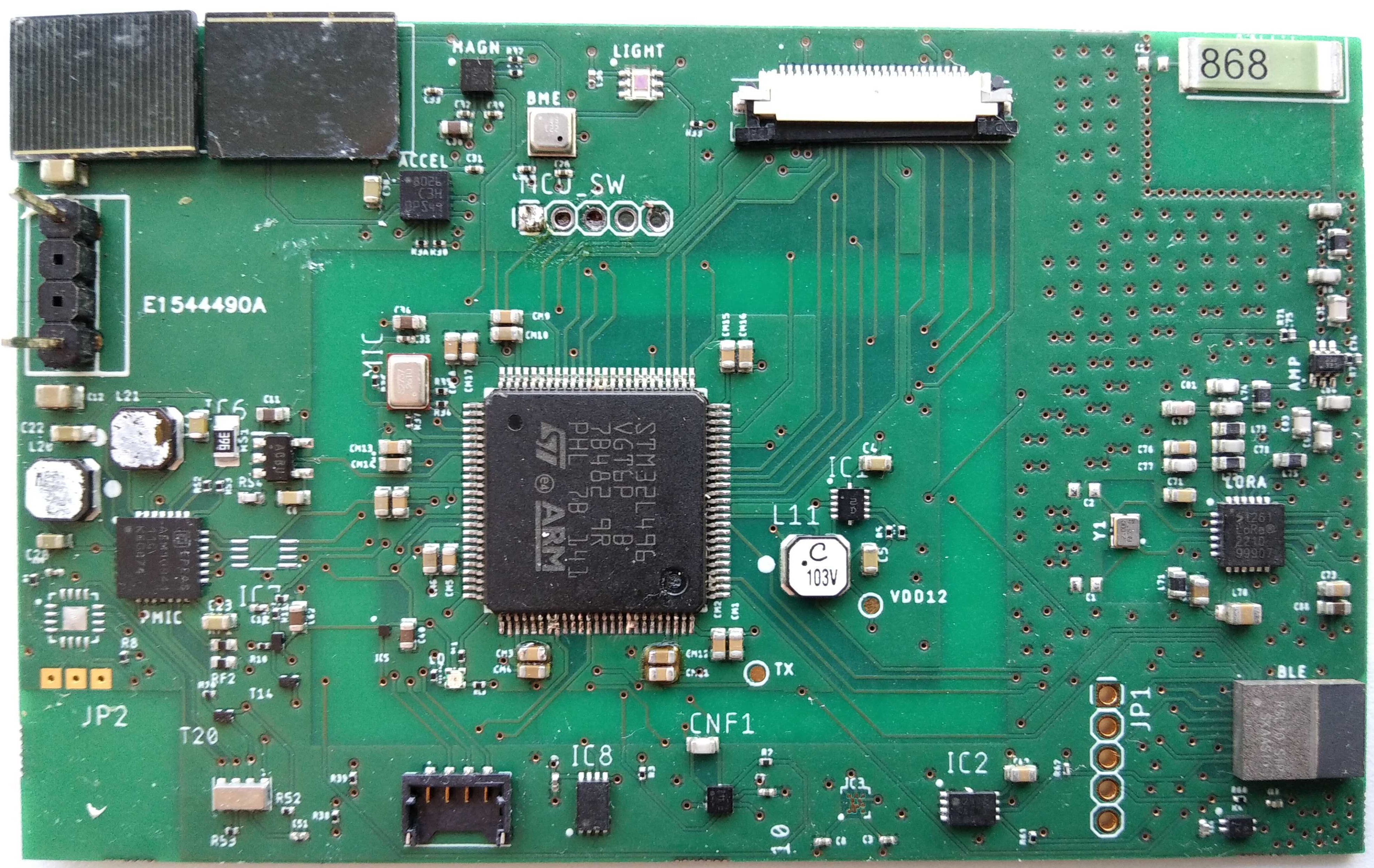} % first figure itself
    \end{minipage}\hfill
    \caption{\gls{ssn} prototype}
    \label{Fig:miniaturized2}
    \Description[Depiction of the proposed system]{The final, miniaturized system.}
\end{figure}

To evaluate the \gls*{dpm} architecture presented in section~\ref{sec:power_management}, a miniaturized, autonomous \gls*{ssn} was designed and developed, powered by \gls*{pv} energy harvesting. The system integrates an $80MHz$ ARM Cortex-M4 \gls*{mcu}, wireless \glspl*{ic} and multiple sensors, powered by a $10mAh$ battery and a $2cm^2$ high-efficiency \gls*{pv} panel. 
This section provides details on the \gls{ssn} implementation, as depicted in Figure~\ref{Fig:Architecture}. A prototype of the printed system can be seen in Figure~\ref{Fig:miniaturized2}.

The architecture of the sensor node can be divided into four interconnected component blocks, the edge intelligence, power management, wireless and sensing blocks.
The edge intelligence block includes the \gls*{mcu} and a non-volatile FRAM for data storage.
The power management block integrates the proposed \gls{dpm} architecture. It includes the \Gls*{pmic}, energy harvester and energy storage, as well as two interfacing components; an \Gls*{rtc} and a capacitive touch sensor.
The wireless block includes \Gls*{ble} and NFC for short-range communication and LoRa for long-range communication. An additional FRAM module is used for the storage of data from the wireless \glspl*{ic}.
Finally, the sensing block includes a CO$_2$ sensor, accelerometer, temperature sensor, magnetometer and light sensor.

% The critical components for the edge intelligence and power management blocks were benchmarked over the state of the art, mainly on their energy consumption characteristics as obtained from their respective datasheet. More information is presented in sections~\ref{sec:edge_block} and~\ref{sec:power_block} below. 
% The selection was also influenced by such factors as dimensions, additional features, or availability at the time of integration.
% For the wireless and sensing blocks, typical low-power, miniaturized components have been utilized, mostly off-the-shelf, as given in table~\ref{tab:Typical}. 
% For the evaluation of the proposed system's energy consumption, as presented in section~\ref{Section:Consumption}, only a subset of these components has been utilized.

% \begin{table*}
%   \caption{Typical system components}
%   \label{tab:Typical}
%   \begin{tabular}{cc|cc}
%     \toprule
%     Model & Type & Model & Type\\
%     \midrule
%     TI OPT3001DNPR~\cite{OPT3001DNPR} & Light sensor &  Knowles SPH0645LM4H~\cite{SPH0645LM4H} & Microphone\\
%     ST LIS3MDLTR~\cite{LIS3MDL} & Magnetometer & NXP NT3H2111W0FHKH~\cite{NT3H2111W0FHK} & NFC\\
%     Semtech SX1261IMLTRT~\cite{SX1261} & LoRa & Onsemi RSL10~\cite{RSL10} & Bluetooth Low Energy\\
%     Bosch BME680~\cite{BME680} & Environmental sensor & ST LIS3DH~\cite{LIS3DH} & Accelerometer \\
%   \bottomrule
% \end{tabular}
% \end{table*}

\subsection{Edge intelligence block}\label{sec:edge_block}
\subsubsection{MCU}
For the \gls*{ssn}, an $100$-pin \gls*{mcu} based on Cortex-M4 is used, stm32l496vg. The \gls*{mcu} has a frequency of $80MHz$, $1MB$ Flash, $25nA$ quiescent current in shutdown mode and $37\mu A/MHz$ energy consumption in run mode when a Switched Mode Power Supply (SMPS) is used. 
In this work, stm32l496vg is supplied by an external SMPS, bypassing the internal regulator of the \gls*{mcu}.

% The newer STM32 U5 \glspl*{mcu}, based on the Cortex-M33 core, have very promising specifications, with a frequency of $160MHz$, up to $4MB$ of flash memory and approximately $18.5\mu A/MHz$ energy consumption and will be considered as future work.

The \gls*{mcu} supports two low-power instructions to transition the system into reduced-power states, Wait-For-Event (WFE) and Wait-For-Interrupt (WFI). WFE wakes-up the \gls*{mcu} when the \textit{event} bit in the System Control Block is set, while WFI keeps the \gls*{mcu} in sleep mode until an interrupt occurs. Since the integrated sensors generate hardware interrupts, the system firmware primarily employs WFI, allowing the MCU to enter low-power states between processing cycles and thus minimize overall energy consumption.

The \gls*{mcu} supports eight low-power operating modes, providing different trade-offs between performance and energy use. \cite{lowpowermodes} presents a comparison between the modes in terms of wake-up source, active components and power consumption for an $1.8V$ input voltage. In run mode, the designed \gls{ssn} supports dynamic voltage scaling to optimize power consumption. The CPU, Flash and SRAM are enabled at a reduced frequency of $2MHz$. In sleep and low-power sleep modes, the CPU halts while peripherals remain active. An interrupt or event can wake-up the CPU. In stop $2$ mode, most of the \gls*{mcu} Vcore domain is put into a lower leakage mode, in order to achieve minimal power consumption while retaining SRAM and register contents.

When executing instructions from Flash memory, wait states may be inserted depending on the CPU clock frequency and internal voltage range. The main regulator output voltage, Vcore, supports two modes that affect the total number of wait states and the total power consumption. Range $1$, or high-performance range, supports a clock frequency of up to $80MHz$ with a minimum Flash access time for reading. Range $2$, or low-power range, has a maximum clock frequency of $26MHz$ and a longer reading time from the Flash memory than Range $1$. The correspondence between wait states and CPU clock frequency is presented in \cite{lowpowermodes}.

% The Flash memory interface of the \gls*{mcu} includes a $256B$ data cache memory with $8$ cache lines of $4 \times 64$ bits each. When data is requested by the CPU, frequently used data lines can be stored in the cache in order to accelerate code execution by enabling a data cache enable bit in the Flash access control register. Moreover, the use of the cache memory for accessing the Flash memory has no effect on the performance of the proposed implementation  when there are no wait states. However, according to \cite{memory}, the cache should lower the power consumption by up to $20\%$, since accesses to the cache require significantly less current compared to accesses to the Flash memory. 

% \begin{table*}
%   \caption{Specifications and electrical characteristics of STM32 \glspl*{mcu}}
%   \label{tab:MCUs}
%   \begin{tabular}{ccccccccl}
%     \toprule
%     Model & Core & Frequency & Flash & RAM & Shutdown & Standby & Stop & Run\\
%     \midrule
%     stm32l010f4 & Cortex-M0+ & $32MHz$ & $16KB$ & $2KB$ & - & $230nA$ & $290nA$ & $76\mu A/MHz$\\
%     stm32l152re & Cortex-M3 & $32MHz$ & $512KB$ & $80KB$ & - & $290nA$ & $560nA$ & $195\mu A/MHz$\\
%     stm32l496vg & Cortex-M4 & $80MHz$ & $1MB$ & $320KB$ & $25nA$ & $108nA$ & $2.57\mu A$ & \begin{tabular}{@{}l}$91\mu A/MHz$  (LDO)\\$37\mu A/MHz$ (SMPS)\end{tabular}\\
%     stm32h753vi & Cortex-M7 & $480MHz$ & $2MB$ & $1MB$ & - & $2.95\mu A$ & $290nA$ & $275\mu A/MHz$\\
%     stm32u5a5vjt & Cortex-M33 & $160MHz$ & $4MB$ & $2.5MB$ & $150nA$ & $195nA$ & $2\mu A$ & $18.5\mu A/MHz$\\
%   \bottomrule
% \end{tabular}
% \end{table*}

\subsubsection{NVM}
The proposed \Gls*{dpm} architecture of this work is built around the concept of completely switching off most system components during inactive periods to eliminate their quiescent current and thus conserve energy. This also includes the system's memory \glspl*{ic}. For this reason non-volatile memories are used to retain data between shut-down/power up cycles; the status of the running application can be stored in the memory before switching off the device, and can be retrieved on subsequent power ups.
For the storage of data between power cycles of this system, the MB85RC64TAPN-G-AMEWE1 FRAM is used. The memory performs write operations at the same speed as read operations, supporting a high speed mode of up to $3.4MHz$. The device can be switched off between transfers to conserve energy and be put into sleep mode which reduces current consumption by stooping the internal regulator circuits. 

% The use of Flash memory is currently prevalent in embedded systems. However, it has significant disadvantages in low-power applications. These include an energy consumption asymmetry between read and write operations, as well as a limited endurance of tens of thousands of access cycles~\cite{meli2022using}. A technology such as a byte-addressable non-volatile memory (NVM), that includes resistive RAM (ReRAM),  magnetoresistive RAM (MRAM) and ferroelectric RAM (FRAM), has superior power and performance characteristics compared to Flash memory~\cite{jayakumar2017energy}. Such characteristics include fast read access, low energy for memory accesses and a high endurance.

% ReRAM, MRAM and FRAM \glspl*{ic} were considered. Since FRAM has a higher endurance than ReRAM ($10^{15}$ vs $10^{5}$) and lower energy requirements per write than MRAM ($2pJ$ vs $120pJ$)~\cite{puglia2019non}, its use was preferred.
 
% Table~\ref{fram_characteristics} summarizes the specifications of the selected memory \gls*{ic}.

To reduce latency, the system includes two such FRAM chips. The first, which is dedicated to the \gls*{mcu}, is connected via an SPI interface. The second that is dedicated to the \Gls*{ble} \gls*{ic}, communicates with the \Gls*{ble} controller using an I$^2$C interface.

\subsection{Power management block}\label{sec:power_block}
% This section details the components that were selected for the power management block of this work. 
% The proposed \gls*{dpm} architecture is detailed in section~\ref{sec:power_management} below. 

\subsubsection{\gls*{pv} harvester and storage}
For the designed \gls{ssn}, two EXL1-1V20-SM cells were chosen; each cell has a surface of $1cm^2$, yielding a total active surface of $2cm^2$. The cells are connected in parallel to increase the output current of the energy harvester.
For energy storage, Powerstream GEB 201212 was used, with a $10mAh$ capacity and dimensions of $2.2 \times 12.5 \times 12.5$mm$^3$.

\subsubsection{\Gls*{pmic}}

\begin{figure}
	\centering		\includegraphics[width=\linewidth]{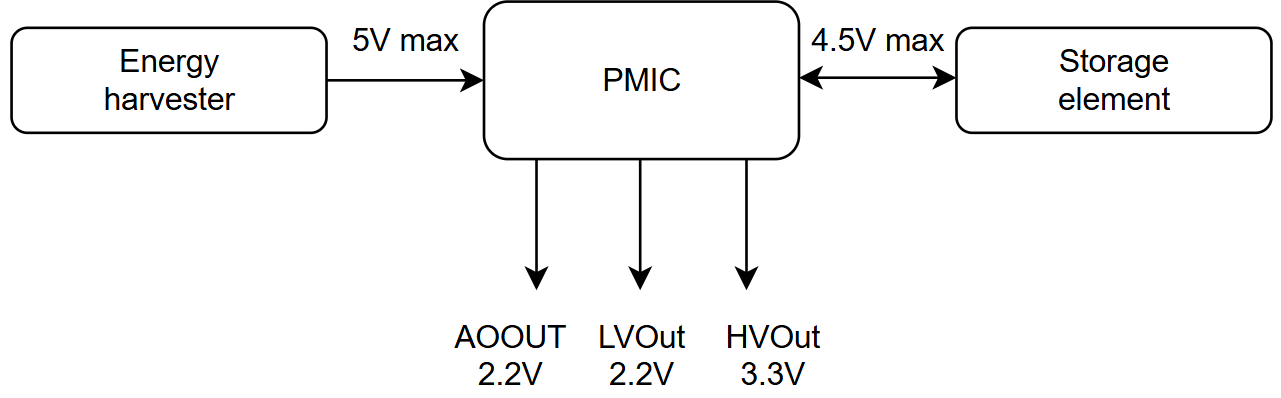}
	\caption{Inputs and outputs of the \Gls*{pmic}}
	\label{Fig:PMIC1}
\end{figure}

For the \gls{ssn}, we employ a \Gls*{pmic} based on AEM10941, offering a cold-start voltage of $300mV$ and a quiescent current of $200nA$.
The \Gls*{pmic} is used for regulating, controlling, and optimizing the power flow from the energy harvester to the battery. It is also used to supply the system's components via three outputs, \gls{aoout}, \gls{lvout} and \gls{hvout}. 
The inputs and outputs of the \Gls*{pmic} can be seen in Figure~\ref{Fig:PMIC1}.

\begin{itemize}
    \item \gls{aoout} is set around a voltage of $2.2V$. It can continuously drive a small current to supply the capacitive sensor and the \Gls*{rtc} even during the sleep mode of the system. The internal power gating switch of the \Gls*{pmic} deactivates this output whenever the energy storage is depleted.
    \item \gls{lvout} is set around a voltage of $2.2V$. It can supply the \gls{mcu}, sensors or any device able to cope with the $2.2V$ supply, through a power gating switch.
    \item \gls{hvout} is a $3.3V$ output of the external DC/DC buck converter allowing to supply any components with higher voltage requirements.
\end{itemize}

\subsubsection{RTC and capacitive touch}
To minimize standby power, our proposed architecture utilizes an external RV-3028-C7 \Gls*{rtc} rather than the \gls*{mcu}-integrated timer. As already discussed, this decoupling allows the \gls*{mcu} and power-demanding peripherals to be switched off during Deep Sleep, while keeping time in the \Gls*{rtc}. 

The RV-3028-C7 \Gls*{rtc} from Micro Crystal is used as part of the power management block and together with the MS8892 capacitive touch sensor are used to wake up the system from the first operating stage. The \Gls*{rtc} has a wide operating voltage range of $1.1V$ to $5.5V$ and a low current consumption of typically $45nA$ at $3V$.
MS8892 has an energy consumption of $65nA$ when an external clock source is provided.

\subsection{Power domains and interfaces}
The \gls{ssn} includes the following domains for the integrated components, corresponding to the outputs of the \Gls*{pmic}.
The \gls{aoout} low-leakage $2.2V$ power domain which is always turned on and is used for the proposed power-management architecture, the \gls{lvout} switched $2.2V$ domain and the optional \gls{hvout} $3.3V$ switched power domain that can be used for components that require higher operating voltage, such as CO$_2$ sensors.
For interfacing, I$^2$C, SPI and I$^2$S are used. In the I$^2$C bus, the \gls*{mcu} operates as the master while the \Gls*{rtc}, sensors and the NFC operate as slaves. I$^2$C is also used for the communication between the \gls*{mcu} and \Gls*{ble} with the respective FRAM \glspl*{ic}. The SPI bus is used for the communication between the LoRa module and the \Gls*{ble} controller. Finally, I$^2$S is used as an interface between the \gls*{mcu} and the microphone.

\section{Performance evaluation}\label{Section:Evaluation}
This section presents a performance evaluation of the proposed hardware-based multi-stage \Gls*{dpm} architecture (section~\ref{sec:power_management}), as integrated into the \gls*{ssn} (section~\ref{Section:Architecture}). We compare the system’s performance against a conventional software-based baseline to quantify the energy gains enabled by full hardware power-gating. 
% Additionally, validation results from a real-world autonomous thermal comfort monitoring case study are presented, confirming sustained operation under ultra-low-light energy-harvesting conditions.

To assess the effectiveness of the proposed \gls*{dpm}, the system’s quiescent current was experimentally measured in Deep Sleep mode, where only the PMIC’s always-on domain, the \gls{rtc} and the capacitive touch sensor remain active.
Measurements were performed using the Nordic Semiconductor Power Profiler Kit II (PPK2), supplying the system at $2.2V$. PPK2 is a standalone unit supporting measurements from $200nA$ up to $1A$, with a resolution between $100nA$ to $1mA$ depending on the range, for a supply voltage of $0.8V$ - $5V$ VCC. 
The quiescent current of the complete system, including the PMIC's own quiescent current, was measured as $452nA$ at $2.2V$, corresponding to the total energy drain of the system while idle. For comparison, the theoretical minimum current, estimated from the device datasheets as the sum of the RTC ($45nA$), touch sensor ($65nA$) and PMIC ($200nA$), is approximately $310nA$.
The difference between theoretical and experimental values can be attributed mainly to leakage currents in passive components and PCB traces as well as to measurement overhead from the PPK2.

To establish a quantitative baseline, we configured the STM32L4 in Stop $2$ mode, representing a state-of-the-art software-based \gls*{dpm}.
The measured current consumption for this configuration was approximately $3\mu A$ at $2.2V$. The comparison highlights the significant reduction in idle current achieved through the proposed power gating approach. Specifically, the hardware-based \Gls*{dpm} achieves a $6.6$x reduction in idle current, effectively eliminating the \gls{mcu}'s quiescent contribution through full domain power-gating.

% Transition Overhead: We measured the energy cost of transitioning between sleep and active states.

% Wake-up (Stage 1 → 2): The transition from a full power-gate to an active MCU (boot-up, retrieve context from FRAM) consumes E_wake = 18.4µJ.

% Shutdown (Stage 2 → 1): The transition from an active MCU to deep sleep (save context to FRAM, clear latch) consumes E_sleep = 8.1µJ.

% The total H-DPM cycle overhead is E_H-DPM_overhead = 26.5µJ.

\subsection{Case study: thermal comfort monitoring}\label{Section:Consumption}

To validate practical viability, we deployed the proposed architecture in a wearable thermal comfort monitoring node for indoor environments.
The system periodically measures environmental temperature and humidity via BME680, computes thermal comfort metrics on the \gls{mcu} and informs a room HVAC via \Gls*{ble} ADV frames using OnSemi RSL10.
For this case study, the edge intelligence and power management blocks were utilized, as discussed in section~\ref{Section:Architecture}.
The firmware was developed on top of ST's Hardware Abstraction Layer (HAL), as generated by STM32CubeMX.

Assuming a fully charged energy storage, the system starts from the first operating stage. Based on periodic, $10$-minute \Gls*{rtc} triggers on 2.2\_SW\_EN, the system proceeds to the second operating stage and the \gls*{mcu}, sensor and \Gls*{ble} \glspl*{ic} are powered. The sensor measures temperature, humidity and atmospheric pressure while the \gls*{mcu} reads previous values from the FRAM. Updated thermal comfort levels are then calculated. The results are transmitted to the HVAC after each measurement in a train of $20$ \Gls*{ble} ADV events. In each event, data is advertised in the $3$ ADV channels. 
% The energy required for \gls*{mcu} processing and data acquisition is low.

Table~\ref{energy_intake} presents the duration of each operating stage and event, the respective energy requirements in $J$ as well as the energy harvested by the energy harvester cells under $200$, $300$ and $500$ lux. As can be seen, the system spends approximately $10$ minutes in the first operating stage, where only the components of the power management block are powered via the 2.2\_LOW\_LEAK rail.

\begin{table}[h]
\caption{Measured energy requirements per event}
\label{energy_intake}
\centering
\begin{tabular}{l l l}
\hline
& Event duration & \begin{tabular}{@{}l}Energy\\required\end{tabular}\\ 
\hline
Data acquisition& $1500ms$ & $1.1mJ$ \\

\gls*{mcu}& $35ms$ &$46.2\mu J$ \\

Always-on domain& \begin{tabular}{@{}l}$10$ mins Deep Sleep +\\$3535ms$ (run mode)\end{tabular}  &$0.6mJ$ \\

System advertising& $2000ms$ & $0.4mJ$ \\

Total per cycle& $3535ms + 10min$ &$2.14 mJ$ \\

% Energy gain & & & $23.9mJ$ & $36.06mJ$ & $70.28mJ$\\
\hline
\end{tabular}
\end{table}

On an \Gls*{rtc} trigger via 2.2\_SW, the system transitions to the second operating stage. There, BME680 acquires temperature, humidity and pressure measurements in approximately $1.5$ seconds and sends them to the \gls*{mcu} via I2C0.
The \gls*{mcu} requires approximately $35$ ms to calculate thermal comfort levels. The results are stored in FRAM. Finally, the results are advertised via \Gls*{ble} to the HVAC controller for $2$ seconds. The \gls*{mcu} then transitions the system back to the first operating stage.
The energy consumption of BME680 for a single measurement of temperature and humidity was measured at $1.1mJ$. The \gls*{mcu} uses an estimated $46.2\mu J$ for the data processing.
The total energy consumption of the power management block for the first and second operating stages was measured at $0.6mJ$. The energy consumption for the ADV events is approximately $0.4mJ$. The total energy cost of the system for a cycle of first and second operating stages is approximately $2.14mJ$.

The energy harvested by the EXL1-1V20 cells under $200$, $300$ and $500$ lux for a complete cycle of operation is measured to $26.04mJ$, $38.2mJ$ and $72.42mJ$ respectively resulting in a net energy gain of $23.9mJ$, $36.06mJ$ and $70.28mJ$.

\section{Conclusions}
\label{Section:Conclusions}
This paper addressed the fundamental limitation of quiescent current drain, which renders traditional software-based \gls{dpm} unfeasible for autonomous systems in energy-scarce, low-light environments. We proposed and implemented a robust, hardware-orchestrated \gls{dpm} architecture for battery-based sensor nodes, based on~\cite{meli2023energy}, that fully power-gates the \gls{mcu} and all non-essential peripherals, entirely eliminating their quiescent current.
A key component of the proposed architecture is the novel hardware latch circuit integrated into the capacitive sensor. The circuit enables the generation of a persistent \textit{enable} signal from a momentary trigger, enabling reliable power-state transitions. This latch circuit, designed by the authors, has been already integrated into the commercial MS8892 \gls{ic}.
Our evaluation demonstrated this architecture achieves a measured sleep-state drain of $452nA$, proving significantly more energy-efficient than traditional software-based sleep modes for battery-based systems. 

% This paper proposed the design and implementation of an autonomous smart sensing embedded system for multi-purpose environmental sensing in smart living and working applications with an emphasis on low-power operation, energy autonomy and miniaturization. It has a form factor of a credit card with a thickness of less than $3mm$.

% A number of crucial components of the system have been specifically designed and optimized for low power consumption and miniaturization. The system's architecture followed a modular approach with the introduction of different functional blocks of components, the edge intelligence, power management, wireless and sensing blocks. The power management block was designed based on a novel power-gating method for the management and conservation of energy. The block utilizes different power domains with multiple power modes of operation that switches off energy to inactive components in order to eliminate their quiescent current.

To validate this architecture, we designed and developed a miniaturized, autonomous \gls{ssn} integrating the proposed \gls{dpm}. This system, was validated in a real-world case study at various illuminance levels. It successfully demonstrated that under $200$, $300$ and $500$ lux for a complete cycle of operation, the system had a net energy gain.

Future work will investigate a more granular, multi-stage power-gating approach. While this work gates the entire compute domain, a hybrid model could keep the MCU's SRAM powered in retention mode while gating the core and peripherals. Finally, we will apply this architecture to higher-performance MCUs, such as the STM32U5 and FRDM-MCXN947, both based on the Cortex-M33 core, to quantify the scaling of energy savings as the quiescent current of high-performance cores continues to rise.

\bibliographystyle{ACM-Reference-Format}
\bibliography{sample-base}

% \appendix









\end{document}